\documentclass[%
 reprint,
 amsmath,amssymb,superscriptaddress,
 aps,
]{revtex4-2}

\usepackage{xcolor}

\usepackage{graphicx}
\usepackage{dcolumn}
\usepackage{bm}
\usepackage{physics}

\begin{document}
\title{Electronic spin susceptibility in metallic strontium titanate}

\author{A. Najev}
\altaffiliation{Present address: Ericsson Nikola Tesla, Krapinska 45, HR-10000 Zagreb, Croatia}
\affiliation{Department of Physics, Faculty of Science, University of Zagreb, Bijeni\v cka 32, HR-10000 Zagreb, Croatia}
\author{N. Somun}
\affiliation{Department of Physics, Faculty of Science, University of Zagreb, Bijeni\v cka 32, HR-10000 Zagreb, Croatia}

\author{M. Spai\'c}
\affiliation{Department of Physics, Faculty of Science, University of Zagreb, Bijeni\v cka 32, HR-10000 Zagreb, Croatia}
\author{I. Khayr}
\affiliation{School of Physics and Astronomy, University of Minnesota, Minneapolis, MN 55455, U.S.A.}
\author{M. Greven}
\affiliation{School of Physics and Astronomy, University of Minnesota, Minneapolis, MN 55455, U.S.A.}
\author{A. Klein}
\affiliation{Department of Physics, Faculty of Natural Sciences, Ariel University, Ariel 40700, Israel}
\author{M. N. Gastiasoro}
\email{maria.ngastiasoro@dipc.org}
\affiliation{Donostia International Physics Center, 20018 Donostia-San Sebastian, Spain}
\author{D. Pelc}
\email{dpelc@phy.hr}
\affiliation{Department of Physics, Faculty of Science, University of Zagreb, Bijeni\v cka 32, HR-10000 Zagreb, Croatia}
\affiliation{School of Physics and Astronomy, University of Minnesota, Minneapolis, MN 55455, U.S.A.}

\begin{abstract} 
Metallic strontium titanate (SrTiO$_3$) is known to have both normal-state and superconducting properties that vary strongly over a wide range of charge carrier densities. This indicates the importance of nonlinear dynamics, and has hindered the development of a clear qualitative description of the observed behaviour.
A major challenge is to understand how the charge carriers themselves evolve with doping and temperature, with possible polaronic effects and evidence of an effective mass that strongly increases with temperature. Here we use $^{47,49}$Ti nuclear magnetic resonance (NMR) to perform a comprehensive study of the electronic spin susceptibility in the dilute metallic state of strontium titanate across the doping-temperature phase diagram. We find a temperature-dependent Knight shift that can be quantitatively understood within a non-degenerate Fermi gas model that fully takes into account the complex band structure of SrTiO$_3$. Our data are consistent with a temperature-independent effective mass, and we show that the behavior of the spin susceptibility is universal in a wide range of temperatures and carrier concentrations. These results provide a microscopic foundation for the understanding of the properties of the unconventional low-density metallic state in strontium titanate and related materials.
\end{abstract}
\maketitle

\section*{Introduction}

Strontium titanate (SrTiO$_3$, STO) is one of the most extensively studied quantum materials and exhibits a host of fascinating properties that remain heavily debated \cite{STOrev,STOSCrev}. Pristine STO is a band insulator that is very close to a ferroelectric instability: it features polar phonon modes that soften dramatically upon cooling, but do not condense to form long-range ferroelectric order \cite{phonon1}. This causes a large low-temperature dielectric permittivity \cite{STOrev} and a strongly nonlinear response to electric fields \cite{nonlinearE}. Moreover, when charge carriers are introduced, the system undergoes an insulator-metal transition at extremely low carrier densities \cite{transport2010,QOPRX,QOPRL} due to the large electronic Bohr radius induced by the high lattice polarizability. Despite extensive experimental and theoretical efforts, the salient features of this low-density metal are not understood. In particular, key questions regarding the nature of the normal-state charge carriers \cite{FL2015,transportPRX,optics} and the electron-phonon interaction in STO and related materials \cite{STOrev} remain open. 

While it is known that the dilute metallic state of STO exhibits sharp Fermi surfaces at low temperatures \cite{QOPRX,QOPRL}, electrical resistivity measurements have revealed an extended Fermi-liquid-like temperature dependence that
cannot be explained by conventional electronic scattering mechanisms \cite{FL2015}. Optical spectroscopy \cite{optics,optics2024} and transport measurements \cite{transportPRX} have provided evidence of highly unusual behavior of the carrier effective mass, which appears to increase significantly with temperature. 
In addition, the low-temperature superconducting state is still controversial, six decades after its discovery \cite{STOSCrev,SC}. It is likely that the coupling of electrons to the soft polar phonon modes plays a pivotal role \cite{TOSCPRL15,STOCaSC,STOSCrev,plastic22,plastic24,energyfluct,STOBaSC}; however, the conventional lowest-order coupling that is linear in the phonon amplitudes is forbidden by symmetry. Several alternative coupling mechanisms have thus been proposed, including two-phonon processes \cite{twophonon0,twophonon1,twophononPRL,twophonon2} and a Rashba-like spin-orbit-assisted interaction \cite{rashba0,rashba1,rashba2,rashba2a,rashba3,rashba4}. Yet the influence of electron-phonon coupling on charge transport is the subject of debate~\cite{bernardi,twophononPRL}, and the extent of polaronic effects on the charge carriers remains unclear. Moreover, a similar low-density metallic state appears in a diverse group of materials systems \cite{STOrev,KTOtransport,BiSO,fullerene}, including thin films and heterostructures \cite{STOLAO,KTOgating1,KTOgating2,KTOEuO}. It is thus of fundamental importance to obtain microscopic understanding.

Here we use nuclear magnetic resonance to gather insight into the nature of the metallic state of STO. While NMR results have been reported for undoped, insulating STO \cite{STONMR,STONMR2}, metallic compositions have 
not been systematically studied with this bulk local probe. 
We use the Knight shift of titanium nuclei to determine the temperature and carrier concentration dependence of the local electronic spin susceptibility and compare the results to detailed calculations that take into account the multi-band structure of STO. Throughout the phase diagram, we find that the behaviour of the spin susceptibility corresponds to a non-degenerate Fermi gas ($T\gg T_F$), with a temperature- and doping-independent effective mass that is only moderately enhanced compared to the non-interacting case. These results clarify the nature of the charge carriers and establish a basis for the understanding of the electronic behavior in STO and related materials with soft polar phonons, such as KTaO$_3$ (KTO).

\section*{Results}

\begin{figure}
\includegraphics[width=85mm]{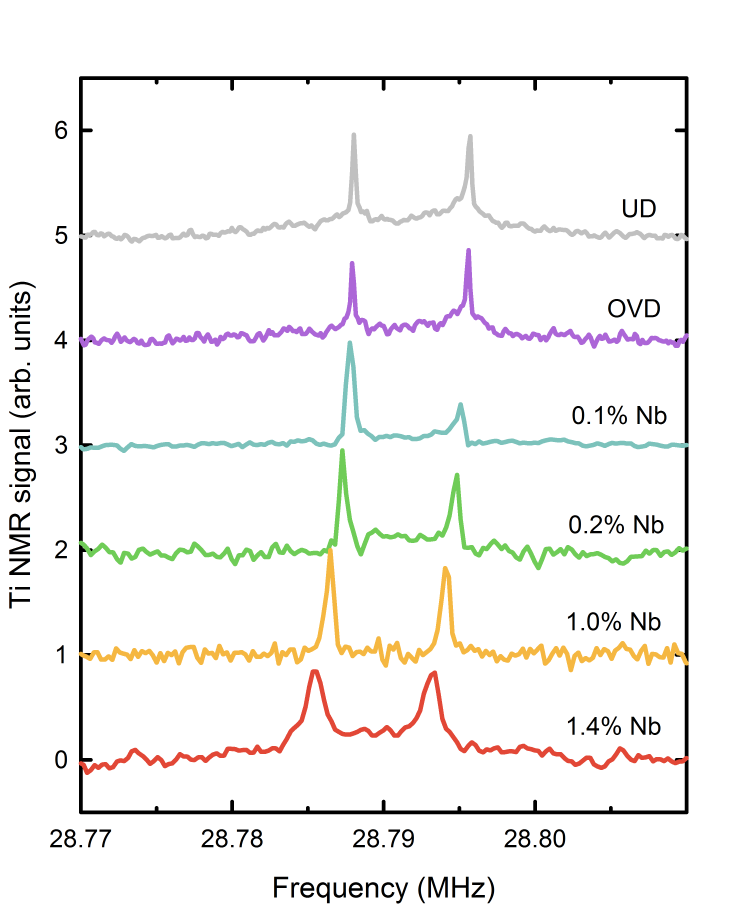}
\caption{\label{fig:spectra} Titanium NMR spectra in STO crystals, measured at 300 K and shifted vertically for clarity. UD and OVD denote undoped and oxygen-vacancy doped samples, respectively, while the niobium concentration (atomic \%) is denoted for Nb-substituted samples. The lower- and higher-frequency peaks correspond to the isotopes $^{47}$Ti and $^{49}$Ti, respectively. The signals are normalized to the values at the low-frequency peaks; the noise levels
differ because of varying sample sizes and electromagnetic penetration depths. }
\end{figure}

Representative ambient-temperature NMR spectra of metallic STO samples are shown in Fig. 1. Both titanium isotopes are visible, and the linewidth increases slightly with doping. The broadening is most likely caused by the coupling of the titanium nuclear quadrupolar moment to local electric field gradients induced by small deviations from the average cubic structure in the vicinity of dopant atoms. Yet the NMR lines remain very narrow up to the highest studied doping level, and the Knight shifts do not exceed a few kHz. This implies that high magnetic field homogeneity and accurate placement of samples in the magnet are essential to obtain reliable shift values. Notably, we do not observe systematic changes of the linewidth at the cubic-to-tetragonal transition at 105~K, which suggests that the associated atomic displacements and disorder due to tetragonal domain walls are too small to affect the titanium nuclei. We therefore ignore quadrupolar effects, and assume that the shift is entirely caused by hyperfine coupling of the nuclei to delocalized charge carriers.

\begin{figure}
\includegraphics[width=70mm]{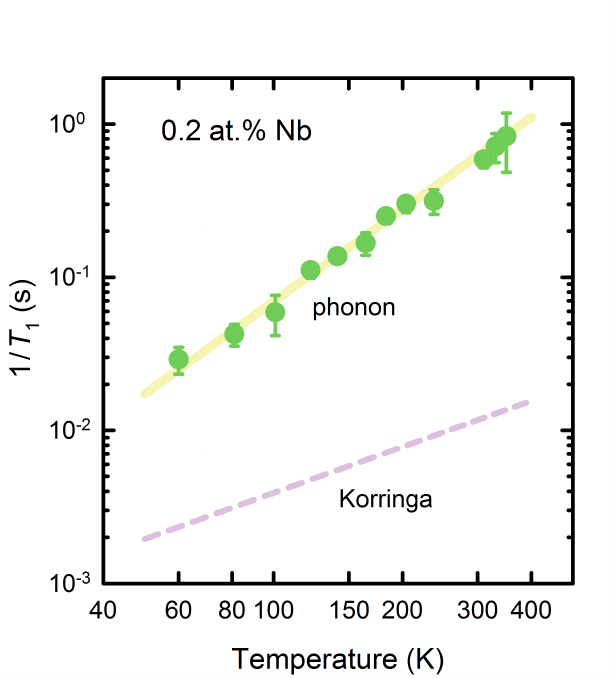}
\caption{\label{fig:T1} Temperature dependence of the titanium nuclear spin-lattice relaxation times $T_1$ for a metallic STO sample substituted with 0.2\% Nb. Similar to insulating KTO \cite{KTONMR}, the behaviour is consistent with $T_1 \sim 1/T^2$ (full line).
The electronic contribution estimated from the low-temperature Knight shift (see Fig. 3) using the standard Korringa relation is plotted as a dashed line. The inconsistency of the latter with experiment indicates that two-phonon Raman scattering 
dominates over electronic relaxation in the entire temperature range.}
\end{figure}

\begin{figure*}
\includegraphics[width=180mm]{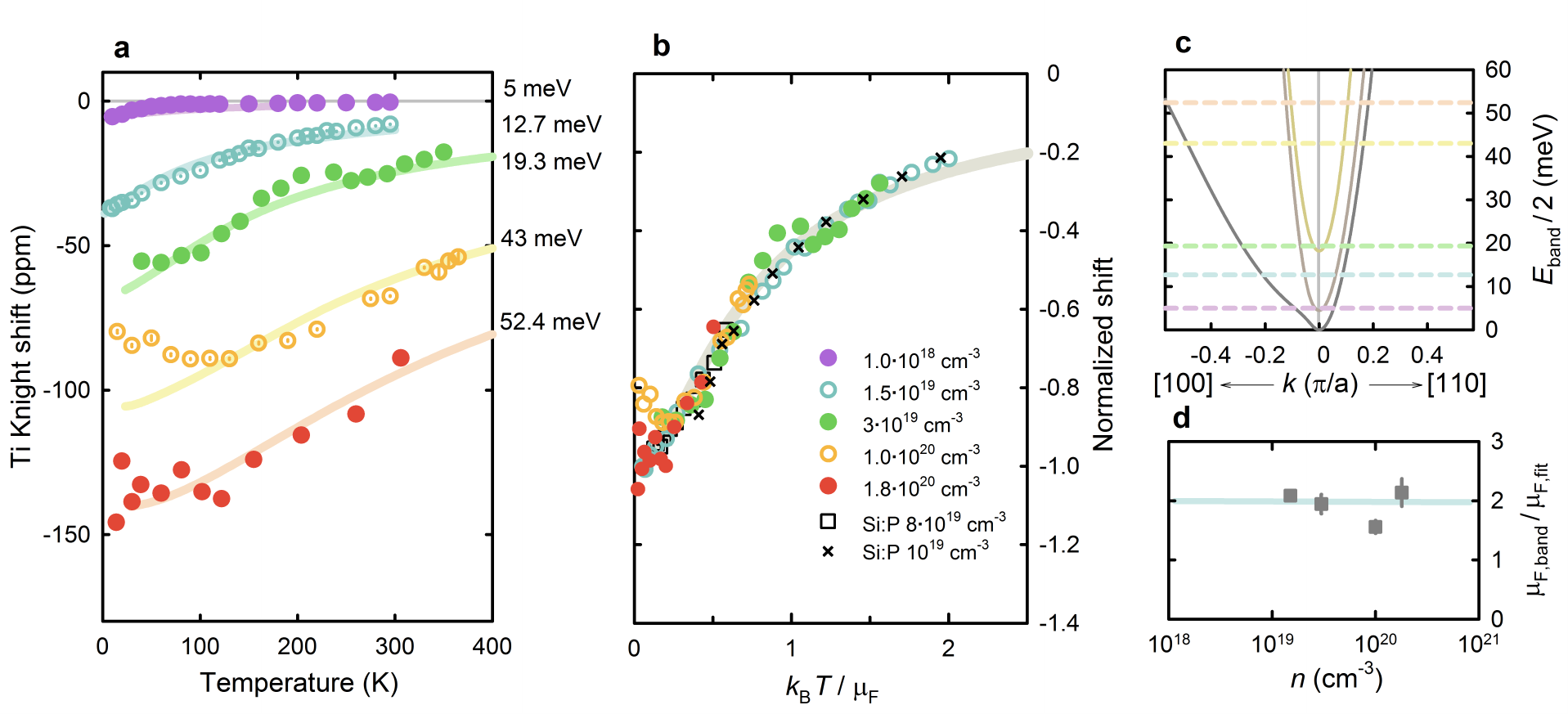}
\caption{\label{fig:shift} (a) Temperature dependence of the titanium Knight shift in metallic STO, for samples doped with oxygen vacancies (OVD) and Nb substitution. The shifts are strongly temperature- and doping-dependent, in agreement with a nondegenerate Fermi gas calculation (full lines). The best-fit zero-temperature chemical potentials $\mu_F$ are indicated for each theoretical curve. (b) Scaling of the Knight shift to the theoretical spin susceptibility, for chemical potentials above $\sim 10$ meV (the Nb-doped samples). A comparison to a conventional low-density metal ($^{31}$P Knight shift in P-doped silicon -- squares and crosses, data from \cite{SiP1,SiP2}) shows similar behaviour. (c) The low-energy electronic dispersion in tetragonal STO obtained from a tight-binding fit to \textit{ab initio} calculations \cite{PRR2023}, along two representative reciprocal space directions. The lowest band, which dominates the spin susceptibility at high chemical potentials, is strongly anisotropic (see also Fig. 4 for Fermi surface plots). The best-fit chemical potentials from (a) are marked as dashed lines. (d) The ratio of the bare-band to best-fit chemical potentials, which is consistent with a roughly twofold, doping- and temperature-independent enhancement of the bare band masses.}
\end{figure*}

In contrast, we find that spin-lattice relaxation predominantly originates from a coupling of the nuclear spin system to phonons. In particular, the relaxation mechanism is consistent with a Raman phonon process \cite{Abragam}, with a characteristic temperature dependence of the form $1/T_1 \sim T^2$, where $T_1$ is the relaxation time and $T$ the temperature (Fig. 2). Although the low signal intensity and long relaxation times preclude more detailed measurements at temperatures below $\sim 50$ K, we note that very similar behavior is also found in undoped KTO \cite{KTONMR}, which supports the absence of electronic effects in $1/T_1$. Phonons can couple dynamically to the titanium nuclear spin via the $^{47,49}$Ti quadrupolar moments, and this coupling is clearly much stronger than the weak hyperfine coupling to conducting electrons. 
A standard Debye-model calculation of the Raman phonon process shows that acoustic phonons yield a $T^2$ relaxation above the Debye temperature \cite{Abragam},  significantly below the Debye temperature, the dependence becomes $\sim T^9$. Since the Debye temperature is $\sim 550$ K in STO \cite{gammaPRM}, the acoustic Raman process cannot explain the observed temperature dependence. Coupling of the nuclear spin system to the zone-center transverse soft optical phonon also yields a $T^2$ contribution for $k_B T \gg \hbar \omega_{TO}$, where $\omega_{TO}$ is the phonon frequency at the $\Gamma$ point \cite{phononT1}. In undoped STO, $\hbar\omega_{TO} \sim 1$~meV~$ \sim 10$ K$/k_B$ at the lowest temperatures, but it increases quite strongly with both temperature and doping. Nonetheless, the Raman process that involves the TO mode seems to be the most likely origin of the observed behavior. Our measurement of a clear $T^2$ dependence indicates that the local nuclear spin relaxation time is insensitive to the electronic renormalization of the transverse optical phonon energy in the 0.2\% Nb substituted material, at least within the studied temperature range.

We also compare the experimental result to the expected electronic contribution to $1/T_1$. To this end,
we use the standard Korringa relation between the spin-lattice relaxation and Knight shift in metals \cite{Abragam}; the low-temperature shift value of $\approx 60$ ppm for the sample with 0.2\% Nb (Fig. 3) yields $1/T_1$ values at least an order of magnitude smaller than what is observed (dashed line in Fig. 2). Given the low carrier density, this is not surprising, but it also implies that spin-lattice relaxation cannot provide meaningful information on the electronic subsystem. We therefore focus on the doping and temperature dependence of the titanium Knight shift to gain insight into the local spin susceptibility.

The measured $^{49}$Ti shifts for all studied samples are shown in Fig. 3a. We first note that the shift relative to pristine STO is always negative; this can be readily understood given the fact that the titanium orbitals that make up the metallic state are of $d$ character. There is thus no spatial overlap between the titanium nuclei and conduction electron wavefunctions, and the usual positive contact hyperfine interaction term is zero. Instead, the electrons interact with the nuclei through an indirect core-polarization process, where the spin polarization of the $d$-like orbitals induces an opposite polarization of inner atomic $s$-orbitals due to the Pauli exclusion principle \cite{core1,Knightshiftrev}. This effect is well known in transition metals, and the core-polarization hyperfine coupling constant for atoms with 3$d$ orbitals, such as titanium, is estimated to be $-125$ kOe/spin \cite{Knightshiftrev}. 

\begin{figure*}
\includegraphics[width=180mm]{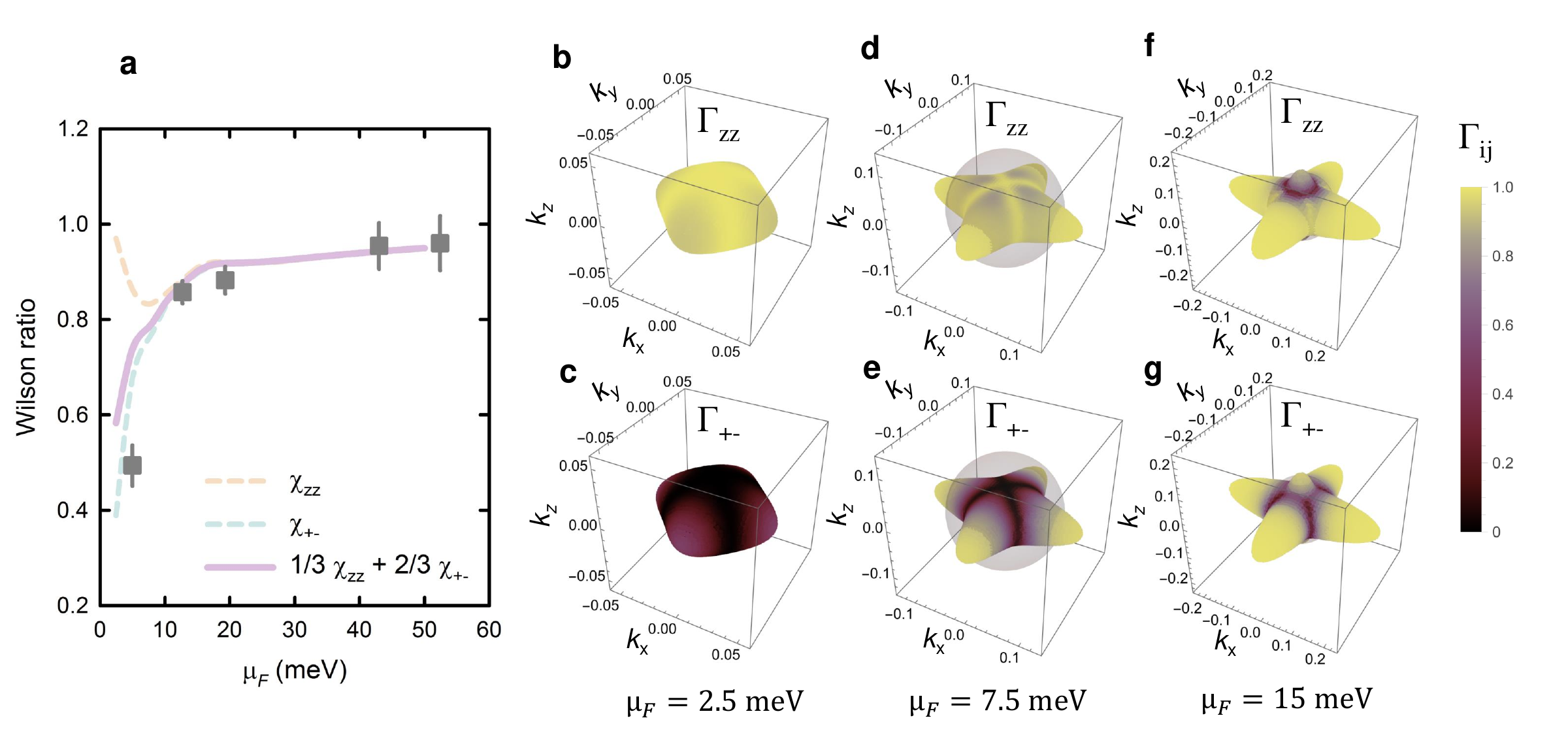}
\caption{\label{fig:gamma} (a) The ratio of magnetic susceptibility obtained from the Knight shift and Sommerfeld coefficient $\gamma$ \cite{gammaPRM} (Wilson ratio) in dependence on the chemical potential at near-zero temperatures. Conventionally, both quantities are proportional to the electronic density of states, which results in a Wilson ratio of one. In metallic STO, the presence of spin-orbit coupling and tetragonal distortions leads to an effective easy-axis anisotropy at low $\mu_F$, which significantly lowers the Wilson ratio for the transverse susceptibility. The dashed lines correspond to the calculated longitudinal and transverse susceptibilities $\chi_{zz}$ and $\chi_{+-}$, respectively, and the full line is an average assuming an equal population of three types of tetragonal domains. The $\mu_F$ of the theoretical model have been divided by a factor of 2 to account for the mass enhancement (see Fig~\ref{fig:shift}). (b-g) the magnitude of the spin susceptibility matrix element $\Gamma_{ij}$ (see Methods) plotted on the Fermi surface of the lowest band, for three values of $\mu_F$. The reciprocal space axes correspond to the cubic structure. A strong suppression of the transverse matrix element is seen at low carrier concentrations (b,c), while the effect becomes progressively weaker as $\mu_F$ increases relative to the tetragonal splitting scale (estimated in (d,e), sphere). }
\end{figure*}

The second important finding is that the Knight shift is strongly temperature-dependent for all studied doping levels, consistent with the spin susceptibility of a non-degenerate Fermi gas ($T\gg T_F$) \cite{Fgaschi}. It is known that the Fermi energies in metallic STO are low, and the qualitative trend of the observed behavior is therefore not surprising. However, our measurements enable us to gain quantitative insight: we compare the data to theoretical spin susceptibilities, which we obtain by taking into account the complex band structure of STO \cite{rashba2}. As established from transport and quantum oscillation measurements, the charge carriers in metallic STO can reside in three distinct bands, which are sequentially filled with increasing doping. Even if interactions are neglected, the electronic spin susceptibility thus has a complex internal structure that includes both interband and intraband contributions (see Methods for details). When the bare-band zero temperature chemical potential $\mu_{F,band}$ is higher than $\sim 20$ meV, the lowest band dominates the behavior of the susceptibility, and the form of the temperature dependence becomes one-band-like, roughly independent of chemical potential. This susceptibility is plotted in Fig. 3b (solid line), with the temperature/energy scale normalized to the chemical potential (horizontal axis), and the susceptibility normalized to the zero-temperature values (vertical axis). The measured Knight shifts for samples with carrier concentration above $\sim 10^{19}$ cm$^{-3}$ show similar scaling, in excellent agreement with the theoretical curve. The scaling is only possible if the electronic effective mass is temperature-independent throughout the studied temperature range, as assumed in the calculation. This is also confirmed through a direct comparison for data for a conventional low-density metal, P-doped silicon \cite{SiP1,SiP2} -- the $^{31}P$ Knight shift agrees very well with the STO data and calculated curve. Moreover, we can compare the chemical potentials that yield the best scaling, $\mu_{F,fit}$, to values calculated from the bare band structure without any interactions, $\mu_{F,band}$. The ratio $\mu_{F,band}/\mu_{F,fit}$ is approximately independent of carrier concentration and close to 2. This is consistent with the enhancement of the effective mass with respect to the bare-band mass that was obtained previously from an analysis of low-temperature specific heat measurements \cite{gammaPRM} and quantum oscillation studies \cite{QO2013,FL2015}. We use the effective mass and best-fit chemical potentials from the scaling analysis to calculate the temperature-dependent Knight shifts, with fits to the individual curves shown in Fig. 3a. We note that the sample with 1\% Nb concentration ($10^{20}$~cm$^{-3}$) shows evidence for a shallow minimum around 100 K; its origin is unknown, and we cannot exclude the possibility of systematic error, e.g., due to slight shifts of the sample in the magnetic field. 

The complex structure of the electronic susceptibility also leads to a strong departure from the usual linear relation between the low-temperature susceptibility and $\gamma$, the electronic Sommerfeld coefficient. The latter was previously determined from specific heat measurements \cite{gammaPRM}, and we can make a direct comparison to the spin susceptibility through the Wilson ratio \cite{Wilson}
\begin{equation}
    R_W = \frac{1}{3}\frac{\pi^2 k_B^2}{\mu_B^2} \frac{\chi}{\gamma},
\end{equation}
where $k_B$ and $\mu_B$ are the Boltzmann constant and Bohr magneton, respectively, and we have assumed an electronic $g$-factor of 2. The Wilson ratio is equal to one for a free electron gas, where both the susceptiblity and Sommerfeld coefficient are proportional to the density of states at the Fermi level. If electron-electron interactions are present, $R_W$ can be greater than one, such as in heavy-fermion materials \cite{heavyfermionWilson} and some iron-based superconductors \cite{pnictideWilson}. Yet in metallic STO we observe a significant \textit{decrease} of the Wilson ratio at low carrier concentrations. This is apparently due to a subtle property of the electronic structure that is captured well by the theoretical calculations (Fig. 4). Indeed,
the calculated spin susceptibility shows strong anisotropy in this regime, even though the tetragonal distortion of the crystal structure is very small. The main reason for both the decrease of the susceptibility and its anisotropy is spin-orbit coupling, which causes the electronic spins to be preferentially aligned with the crystallographic $c$-axis, i.e., the tetragonal distortion induces an effective easy-axis anisotropy (Fig. 4b-c). Above $\mu_F \sim 10$ meV, where the lowest electronic band begins to dominate the response, the effects of spin-orbit coupling and anisotropy become much smaller, as demonstrated in Fig. 4d,e, and the Wilson ratio approaches one. We note that the best agreement between experimental and theoretical Wilson ratio is obtained for a hyperfine coupling constant of -80 kOe/spin, in reasonable agreement with the 3$d$ core polarization value of -125 kOe/spin \cite{core1}. 
The decrease of the Wilson ratio should be even more pronounced in doped KTaO$_3$, where the effects of spin-orbit coupling are expected to be much larger, leading to significantly reduced electronic magnetic moments due to the entangled orbital and spin degrees of freedom~\cite{ahadi}.

\section*{Discussion}

Our results have several important implications for the physics of metallic STO, and dilute metals more generally. First, we show that the behavior of the electronic subsystem is quantitatively consistent with a nondegenerate Fermi liquid in a wide doping and temperature range. This central result implies that quasiparticles are well defined, and that temperature only affects their energy distribution in accordance with conventional Fermi-Dirac statistics. The robustness of the susceptibility scaling provides strong evidence for a temperature-independent effective mass, with a moderate mass enhancement that agrees with previous low-temperature studies \cite{gammaPRM}. The simple observed behavior provides a microscopic underpinning for theories, both to explain superconductivity, and to explain the unusual transport properties of the dilute metal. In particular, our results validate the central assumptions of a model based on two-phonon scattering that explains the persistence of Fermi-liquid-like $T^2$-resistivity at temperatures well beyond the Fermi temperature \cite{twophononPRL}. Yet similar behavior of the resistivity has been found in other dilute metals without soft phonons \cite{BiSO}, and alternative scenarios have been put forward \cite{diffusivity}. The spin susceptibility is a thermodynamic quantity and thus not sensitive to the scattering mechanism, but it does provide an accurate independent estimate of the chemical potential, which we show is in good agreement with the band structure once the mass enhancement is taken into account. Most importantly, our measurements are not limited to the degenerate Fermi liquid regime, and probe the electronic parameters at temperatures comparable to the Fermi energy.

The best-fit chemical potentials (Fig. 3) are also in fairly good agreement with values obtained using other experimental techniques, especially at carrier concentrations below $10^{20}$ cm$^{-3}$. It is known from quantum oscillation studies \cite{QOPRL,FL2015} that the chemical potentials for samples doped with 0.1\% and 0.2\% Nb lie below and above the bottom of the third band, respectively, as also obtained from the Knight shift scaling (Fig. 3c). Furthermore, we can compare the best-fit values to a simple estimate of the chemical potential within the free-electron approximation, i.e. assuming that the band structure can be approximated by a single isotropic parabolic band. At high carrier concentrations, the Sommerfeld coefficient then yields an effective mass of $\sim 4 m_e$, and the chemical potential for the sample with carrier density $1.8 \times 10^{20}$~cm$^{-3}$ can be estimated to be around 30 meV. This is somewhat below our best-fit value of 52.4~meV, and a possible reason for the difference might be the strong anisotropy of the realistic band dispersion. As seen in Fig. 3c and Fig. 4, the Fermi surface for the outermost band includes regions where the curvature changes from convex to concave, with an associated divergence of the density of states. This should substantially affect the Fermi energy calculation for a given carrier density. The discrepancy illustrates the importance of taking into account the realistic band structure, even in the regime where one band dominates the response.

The finding of a temperature-independent effective mass invites a reconsideration of the interpretation of recent charge transport experiments, which has suggested a strong increase of the mass with increasing temperature and a crossing of the Mott-Ioffe-Regel limit \cite{transportPRX}. Evidence for a significant mass enhancement has also been found from the analysis of optical conductivity data \cite{FL2011,optics2024}. We stress that NMR provides direct insight into the microscopic spin susceptibility, and the analysis of this data does not rely on any assumptions on the scattering mechanisms and their energy dependence. Moreover, we explicitly take into account the complex band structure of STO; the latter might affect the transport coefficients in a nontrivial way, especially due to interband terms. Nevertheless, it would be interesting to expand the NMR experiments into the high-temperature regime, where the electronic mean-free-path becomes very short, while the mobility does not show signs of saturation. More generally, our results show that NMR can be a valuable probe of the electronic subsystem in dilute metals, that can be used to study even the small electronic susceptibilities in such materials in a wide range of temperatures and carrier concentrations.

\section*{Methods}

\textit{Samples and NMR.} We have performed $^{47,49}$Ti NMR measurements on commercial (MTI Corp.) oriented single crystals of STO, doped with both oxygen vacancies (OVD) and niobium substitution. The vacancy doping was induced by annealing a pristine STO crystal in high vacuum ($10^{-5}$ mbar) at 900$^{o}$C for two hours, similar to previous work \cite{FL2015}. Charge carrier concentrations were obtained from Hall number measurements or estimated from the niobium concentration, which is accurate to $\sim 10$\%. The NMR spectra were recorded with a conventional spin echo pulse sequence, using Tecmag spectrometers and a high-homogeneity 12~T superconducting magnet (Oxford). The field was always oriented along the crystallographic $\langle 1 0 0 \rangle$ direction. All NMR lineshifts are referenced to undoped STO, and we focus on the $^{49}$Ti isotope due to its slightly larger abundance. Because of small signal levels and slow relaxation rates, spin-lattice relaxation was measured with a single-shot pulse sequence, where the nuclear magnetization is repeatedly probed during a single relaxation process \cite{nutsandbolts}. 

\textit{Theory.} The static noninteracting spin susceptibility is given by
\begin{equation}
\label{eq:chi}
    \chi_{ij}(\mathbf{q}) = -g^2 \mu_B^2 \sum_{m,m'}\sum_{\mathbf{k}} \Gamma_{ij}^{m,m'} \frac{f_m(\mathbf{k})-f_{m'}(\mathbf{k+q})}{\varepsilon_m(\mathbf{k})-\varepsilon_{m'}(\mathbf{k+q})+i\gamma}
\end{equation}
where $g = 2$ and $\mu_B$ are the electron g-factor and Bohr magneton, respectively, $m$ and $m'$ are the band indices, $f_m(\mathbf{k})$ the Fermi-Dirac function, $\varepsilon_m(\mathbf{k})$ the electronic dispersion, $\gamma$ a small broadening, and the matrix element is
\begin{equation}
    \Gamma_{ij}^{m,m'}=\langle \mathbf{k},m | S_i | \mathbf{k+q},m'\rangle \langle \mathbf{k+q},m' | S_j | \mathbf{k},m\rangle
\end{equation}
where $S_i$ are spin-1/2 operators, and $i,j = x,y,z$. The matrix elements for the lowest band $m=m'=1$ are shown in Fig~\ref{fig:gamma}b-g. We compute the $\mathbf{q} \xrightarrow{} 0$ limit of the transverse susceptibility $\chi_\pm(\bm{q}\rightarrow \bm 0)$ with spins in the $xy$ plane and longitudinal susceptibility $\chi_{zz}(\bm{q}\rightarrow \bm 0)$ with spins along the tetragonal $z$-axis. To evaluate Eq.~\eqref{eq:chi} we use a tight-binding parametrization of the relativistic STO band structure obtained from first-principles calculations~\cite{rashba2,PRR2023}. The sphere in panels \ref{fig:gamma}d-e shows the tetragonal energy scale of this model, which is given by the split of the $d_{yz}/d_{zx}$ and $d_{xy}$ orbitals obtained from \emph{ab initio} due to the tetragonal crystal field. The susceptibility is calculated on a grid of energy/temperature and chemical potential values, and we use polynomial interpolation to fit to the data. The Sommerfeld coefficient in Fig. 4 is computed in a similar way, but with the matrix elements set to 1. 

\section*{Acknowledgements}

We thank Ivan Jakovac, Chris Leighton, Boris Shklovskii, Turan Birol, Beno\^{i}t Fauqu\'{e} and Kamran Behnia for discussions and comments. The work at the University of Zagreb was supported by the Croatian Science Foundation through Grant No. UIP-2020-02-9494, using equipment funded in part through project CeNIKS co-financed by the Croatian Government and the European Union through the European Regional Development Fund - Competitiveness and Cohesion Operational Programme (Grant No. KK.01.1.1.02.0013). The work at the University of Minnesota was funded by the US Department of Energy through the University of Minnesota Center for Quantum Materials, under Grant No. DE-SC-0016371. M.N.G is supported by the Ramon y Cajal Fellowship RYC2021-031639-I. A.K. acknowledges support by the Israel Science Foundation (ISF) and the Israeli Directorate for Defense Research and Development (DDR\&D) under grant No. 3467/21, and by the  United States - Israel Binational Science Foundation 
(BSF), grant No. 2022242.

\newpage
\bibliography{main}
\end{document}